\documentclass[aps,pra, article, amsmath, amssymb, showpacs, superscriptaddress]{revtex4}

\usepackage{graphicx}% Include figure files
\usepackage{bm}% bold math
\begin{document}

\title{Mass Defect Effects in Atomic Clocks}

\author{V. I. Yudin}
\email{viyudin@mail.ru}
\affiliation{Novosibirsk State University, ul. Pirogova 2, Novosibirsk, 630090, Russia}
\affiliation{Institute of Laser Physics SB RAS, pr. Akademika Lavrent'eva 13/3, Novosibirsk, 630090, Russia}
\affiliation{Novosibirsk State Technical University, pr. Karla Marksa 20, Novosibirsk, 630073, Russia}
\author{A. V. Taichenachev}
\affiliation{Novosibirsk State University, ul. Pirogova 2,
Novosibirsk, 630090, Russia}
\affiliation{Institute of Laser Physics SB RAS, pr. Akademika
Lavrent'eva 13/3, Novosibirsk, 630090, Russia}
%

%\date{\today}

\begin{abstract}
We consider some implications of the mass defect on the frequency
of atomic transitions. We have found that some well-known
frequency shifts (gravitational shift and motion-induced shifts such as: quadratic Doppler and micromotion
shifts) can be interpreted as consequences of the mass defect in quantum atomic physics,
i.e., without the need for the concept of time dilation used in
special and general relativity theories. Moreover, we show that
the inclusion of the mass defect leads to previously unknown
shifts for clocks based on trapped ions.
\end{abstract}


\maketitle

\section*{Introduction}
At the present time, atomic clocks are most precise scientific devices. The principle of operation of these quantum instruments is based on modern methods of laser physics and high-precision spectroscopy. In this way, the unprecedented value of fractional instability and uncertainty at the level of 10$^{-18}$ has already been achieved with the goal of 10$^{-19}$ on the horizon \cite{Schioppo_2017}. Frequency measurements at such a level could have a huge influence on further developments in fundamental and applied physics. In particular, we can foresee tests of quantum electrodynamics and cosmological models, searches for drifts of the fundamental constants, new types of chronometric geodesy, and so on (see, for example, review \cite{Ludlow_2015}). However, this level of experimental accuracy requires a comparable level of theoretical support, which would account for systematic frequency shifts of atomic transitions due to different physical effects. Thus, modern atomic clocks are also at the point of interweaving different areas of theoretical physics.

In this Letter we develop the mass defect concept with respect to
atomic clocks.  Historically, considerations of the mass defect
have been connected with nuclear physics, first of all, where the mass defect
explains the huge energy emitted due to different nuclear
reactions. However, a quite unexpected result is that this effect
has a direct relation to frequency standards, where it leads to
the different shifts in the frequencies of atomic transitions.

The main idea of our approach is the following. Let us consider an arbitrary atomic transition between two states $|{\rm g}\rangle$ and $|{\rm e}\rangle$ with unperturbed frequency $\omega^{}_0=(E^{(0)}_{\rm e}-E^{(0)}_{\rm g})/\hbar$, where $E^{(0)}_{\rm g}$ and $E^{(0)}_{\rm e}$ are unperturbed energies of the corresponding states (see Fig.~\ref{Fig1}). Using Einstein's famous formula, $E=Mc^2$, which links the mass $M$ and energy $E$ of a particle ($c$ is the speed of light), we can find the rest masses of our particle,  $M_{\rm g}$ and $M_{\rm e}$, for the states $|{\rm g}\rangle$ and $|{\rm e}\rangle$, respectively: $E^{(0)}_{\rm g}=M_{\rm g}c^2$ and $E^{(0)}_{\rm e}=M_{\rm e}c^2$. The fact that $M_{\rm g}\neq M_{\rm e}$ is the essence of the so-called mass defect (or mass difference). In our case, the connection between $M_{\rm g}$ and $M_{\rm e}$ is the following:
\begin{equation}\label{M_e}
M_{\rm  e}c^2=M_{\rm g}c^2+\hbar \omega^{}_0\quad \Rightarrow\quad M_{\rm e}=M_{\rm g}+\frac{\hbar\omega^{}_0}{c^2}\,.
\end{equation}
Note that several years ago an idea of the direct measurement of the mass difference $\delta M=M_{\rm  e}-M_{\rm  g}$ under emission of gamma rays (from nuclei) has been actively discussed (e.g., see Ref.~\cite{Pritchard_2002}).

In this Letter we show that the relationship (\ref{M_e}) allows us to
reinterpret some well-known systematic frequency shifts (such as
the so-called time dilation effects
\cite{Einstein_1905,Einstein_1916,Chou_Sc_2010}) in atomic clocks. Moreover, our
approach actually predicts some shifts previously
unconsidered, to our knowledge, in the scientific literature.

\section{Gravitational shift}
As a first example, let us show, for generality, how the mass defect allows us to formulate methodologically simplest explanation of the gravitational redshift, even with a classical description of the gravitational field (as classical Newtonian potential $\varphi$). Indeed, because the potential energy of a particle in a classical gravitational field is equal to the product $M\varphi$ (where $\varphi <0$), we can write the energy of $j$-th state $E_j(\varphi)$ as:
\begin{equation}\label{E_G}
E_j(\varphi)=M_{j\,}c^2+M_{j\,}\varphi=M_{j\,}c^2\left(1+\frac{\varphi}{c^2}\right),\quad
(j={\rm e,g}).
\end{equation}
Here we have used an equality of gravitational and inertial masses, $M_{\rm grav}=M_{\rm in}$.
Using Eqs.~(\ref{E_G}) and (\ref{M_e}), we find the frequency of the transition $|{\rm g}\rangle\leftrightarrow|{\rm e}\rangle$ in the gravitational field:
\begin{equation}\label{omega_G}
\omega=\frac{E_{\rm e}(\varphi)-E_{\rm g}(\varphi)}{\hbar}=\omega^{}_0\left(1+\frac{\varphi}{c^2}\right).
\end{equation}
This expression coincides (to leading order) with a well-known result based on general relativity theory \cite{Einstein_1916}:
\begin{equation}\label{rel_omega_G}
\omega=\omega^{}_0\sqrt{1+\frac{2\varphi}{c^2}}\approx\omega^{}_0\left(1+\frac{\varphi}{c^2}\right),
\end{equation}
in the case of $|\varphi/c^2|\ll 1$.

Thus, the combination of special relativity ($E=Mc^2$) and quantum
mechanics (definition of the frequency of atomic transition)
leads to a formally noncontradictory explanation of the
gravitational shift (\ref{omega_G}), which describes different
experiments (e.g., see
\cite{Chou_Sc_2010,Katori_2016,Pound_1960}) with atomic clocks in a spatially nonuniform
gravitational potential $\varphi({\bf r})$. Indeed, using Eq.~(\ref{omega_G}) we can write well-know expression (in the case of $|\varphi/c^2|\ll 1$):
\begin{equation}\label{frac_w}
\frac{\omega({\bf r}_1)-\omega({\bf r}_2)}{\omega({\bf r}_1)}=\frac{[\varphi({\bf r}_1)-\varphi({\bf r}_2)]/c^2}{1+\varphi({\bf r}_1)/c^2}\approx \frac{\varphi({\bf r}_1)-\varphi({\bf r}_2)}{c^2}\,,
\end{equation}
which connects the fractional shift $\Delta\omega/\omega$ with the difference of gravitational potentials $[\varphi({\bf r}_1)-\varphi({\bf r}_2)]$ between emitter (coordinate ${\bf r}_2$) and observer (coordinate ${\bf r}_1$).

Note that Eq.~(\ref{omega_G}) was derived without including the
gravitational time dilation, which is taken as a basis of
Einstein's theory of the general relativity \cite{Einstein_1916}.
Therefore, the mass defect approach (see also in Ref.~\cite{Wolf_2016}) can be considered as
``quasi-classical'' explanation of the gravitational shift.

\section{Motion-induced shifts for a free atom}
As a second example, we will perform an exact
quantum-relativistic calculation of the frequency of transition
under one-photon absorption and emission by a free-moving atom. We
assume that the energy and momentum of the atom+photon system are
conserved. Considering the law of energy conservation for
absorption/emission of a photon with frequency $\omega$, the
following relativistic relationship results:
\begin{equation}\label{E_rel}
\hbar\omega=\sqrt{c^2{\bf
p}_{\rm e}^2+M_{\rm e}^2c^4}-\sqrt{c^2{\bf
p}_{\rm g}^2+M_{\rm g}^2c^4}\,,
\end{equation}
where ${\bf p}_{\rm e}$ and ${\bf p}_{\rm g}$ are the momenta of
an atom in the internal states $|{\rm e}\rangle$ and
$|{\rm g}\rangle$, respectively.

Let us first consider the absorption of a photon with momentum
$\hbar {\bf k}={\bf n}\hbar\omega /c$, where the unit vector ${\bf
n}={\bf k}/|{\bf k}|$ is directed along the wave vector ${\bf k}$.
We assume that initially the atom is in the lower state
$|{\rm g}\rangle$ and has the momentum ${\bf p}_{\rm g}$. Then,
in accordance with conservation of momentum, we find the atomic
momentum in the excited state $|{\rm e}\rangle$ after absorption
of photon to be: ${\bf p}_{\rm e}={\bf p}_{\rm g}+{\bf
n}\hbar\omega /c$. In this case, Eq.~(\ref{E_rel}) has the
following form:
\begin{equation}\label{E_rel_abs}
\hbar\omega=\sqrt{c^2({\bf p}_{\rm g}+{\bf n}\hbar\omega
/c)^2+M_{\rm e}^2c^4}-\sqrt{c^2{\bf
p}_{\rm g}^2+M_{\rm g}^2c^4}\,,
\end{equation}
which should be considered as an equation for the unknown
frequency $\omega$. Taking into account (\ref{M_e}), the exact
solution of Eq.~(\ref{E_rel_abs}) can be written as:
\begin{equation}\label{omega_abs_p}
\omega=\frac{\omega_0+\omega^{({\rm rec})}_{\rm g}}{\sqrt{1+{\bf
p}^2_{\rm g}/(M_{\rm g}c)^2}-({\bf n}\cdot {\bf
p}_{\rm g})/(M_{\rm g}c)}\,,
\end{equation}
where $\omega^{({\rm rec})}_{\rm g}=\hbar
\omega^2_0/(2M_{\rm g}c^2)$ is the recoil frequency, which
is the correction associated with the so-called recoil effect. For
atomic transitions of optical range ($\omega_0/2\pi\sim
10^{14}$--$10^{15}$~Hz), we have the following estimate:
$\omega^{({\rm rec})}/2\pi\sim 1$--$1000$~kHz. Using the
relativistic relationship between momentum ${\bf p}_{\rm g}$ and
velocity ${\bf v}_{\rm g}$ (for the lower state
$|{\rm g}\rangle$): ${\bf p}_{\rm g}=M_{\rm g}{\bf
v}_{\rm g}/\sqrt{1-{\bf v}_{\rm g}^2/c^2}$, formula
(\ref{omega_abs_p}) can be rewritten:
\begin{equation}\label{omega_abs_v}
\omega=\left(\omega_0+\omega^{({\rm rec})}_{\rm g}\right)\frac{\sqrt{1-{\bf
v}_{\rm g}^2/c^2}}{1-({\bf n}\cdot {\bf v}_{\rm g})/c}\,.
\end{equation}
Let us calculate now the frequency of a photon emitted along the
line of ${\bf n}$ by the moving atom, which was initially in the
upper level $|{\rm e}\rangle$ and had momentum ${\bf
p}_{\rm e}$. In this case, the momentum of the atom in the lower
level $|{\rm g}\rangle$ after emission of the photon becomes ${\bf
p}_{\rm g}={\bf p}_{\rm e}-{\bf n}\hbar\omega /c$. Solving
Eq.~(\ref{E_rel}), we obtain the following result:
\begin{equation}\label{omega_rad}
\omega =\frac{\omega_0-\omega^{({\rm rec})}_{\rm e}}{\sqrt{1+{\bf p}^2_{\rm e}/(M_{\rm e}c)^2}-({\bf n}\cdot {\bf p}_{\rm e})/(M_{\rm e}c)}
=\left(\omega_0-\omega^{({\rm rec})}_{\rm e}\right)\frac{\sqrt{1-{\bf
v}_{\rm e}^2/c^2}}{1-({\bf n}\cdot {\bf v}_{\rm e})/c},
\end{equation}
where $\omega^{({\rm rec})}_{\rm e}=\hbar
\omega^2_0/(2M_{\rm e}c^2)$ is the correction due to the recoil,
and the velocity ${\bf v}_{\rm e}$ and momentum ${\bf
p}_{\rm e}$ are connected by the relationship: ${\bf
p}_{\rm e}=M_{\rm e}{\bf v}_{\rm e}/\sqrt{1-{\bf
v}_{\rm e}^2/c^2}$.

For comparison, let us consider the absorbing/emitting classical
oscillator with eigenfrequency $\omega_0$, which moves with
velocity ${\bf v}$ relative to laboratory system of coordinates.
In this case, using the Lorentz transformations, we obtain the
following well-known expression for the frequency of the
electromagnetic wave under absorption/emission along the line
${\bf n}$:
\begin{equation}\label{omega_rel}
\omega=\omega_0\frac{\sqrt{1-{\bf v}^2/c^2}}{1-({\bf n}\cdot {\bf
v})/c}\,.
\end{equation}
Comparing Eq.~(\ref{omega_rel}) with formulas (\ref{omega_abs_v})
and (\ref{omega_rad}), we see that the quantum-relativistic
calculation significantly differs from the
classically-relativistic version. Formally this can be seen as a
renormalization of the eigenfrequency $\omega_0$:
$\omega_0\rightarrow \omega_0+\omega^{({\rm rec})}_{\rm g}$ in
the case of absorption [see Eq.~(\ref{omega_abs_v})], and
$\omega_0\rightarrow \omega_0-\omega^{({\rm rec})}_{\rm e}$ in
the case of emission [see Eq.~(\ref{omega_rad})].

Note that the interrelation of the second-order Doppler shift and
mass defect in the context of free-moving object was previously
shown for M$\ddot{{\rm o}}$ssbauer effect in
Ref.~\cite{Josephson_1960,Dehn_1970}, where the usual Doppler formula
(\ref{omega_rel}) was derived as a consequence of conservation of
energy and momentum instead of the phase invariance. However,
the derivation in Ref.~\cite{Dehn_1970} is not quite suitable for atomic clocks, because it did not include the nonzero recoil shift [see $\omega^{({\rm rec})}_{\rm g,e}$ in
the exact expressions (\ref{omega_abs_v}) and (\ref{omega_rad})], which is very
important for frequency standards based on free atoms. Our formulas (\ref{omega_abs_v}) and
(\ref{omega_rad}) will coincide with Eq.~(\ref{omega_rel}) in the
limit $M_{\rm g,e}\rightarrow\infty$, when
$\omega^{({\rm rec})}_{\rm g,e}\rightarrow 0$.

Thus, we have above shown that for one-photon transition the Doppler formula (\ref{omega_rel}), based on the Lorentz transformations and time delay concept, does not coincide with exact quantum-relativistic expressions (\ref{omega_abs_p}), (\ref{omega_abs_v}) and (\ref{omega_rad}), based on the mass defect concept. It occurs because the velocities in the ground and exited states are not equal to each other, ${\bf v}_{\rm g}\neq{\bf v}_{\rm e}$, due to the recoil effect, first of all. Consequently, a unique inertial reference frame, which can be unambiguously associated with the moving atom, does not exist. Moreover, let us show that even in the total absence of the recoil effect for two-photon transition, formed by two counter propagating waves with the same frequency $\omega\approx\omega_0/2$ (i.e., standing wave), the well-known formula based on the Lorentz transformations:
\begin{equation}\label{omega_2_ph}
\omega=\frac{\omega_0}{2}\sqrt{1-{\bf v}^2/c^2}\,,
\end{equation}
also is not exact. Indeed, in the absence of the recoil effect, momentums in the ground and exited states are the same, ${\bf p}_{\rm g}={\bf p}_{\rm e}={\bf p}$. Nevertheless, using relativistic expressions:
\begin{equation}\label{V_j}
{{\bf p}}=\frac{M_j{{\bf v}}_j}{\sqrt{1-{{\bf v}}_j^2/c^2}}\;\;\Rightarrow \;\;{{\bf v}}_j=\frac{{{\bf p}}/M_j}{\sqrt{1+{{\bf p}}^2/(M_j\,c)^2}}\,,\;\; (j={\rm g,e})\,,
\end{equation}
we see that the velocities are not already the same, ${\bf v}_{\rm g}\neq{\bf v}_{\rm e}$, due to the mass defect ($M_{\rm g}\neq M_{\rm e}$). Therefore, the use of Eq.~(\ref{omega_2_ph}) leads to the ambiguity:
\begin{equation}\label{omega_amb}
\omega(1)=\frac{\omega_0}{2\sqrt{1+{{\bf p}}^2/(M_{\rm g}\,c)^2}}\,,\;\;\;\; \omega(2)=\frac{\omega_0}{2\sqrt{1+{{\bf p}}^2/(M_{\rm e}\,c)^2}}\,.
\end{equation}
However, the exact result can be expressed as the following:
\begin{equation}\label{E_2_ph}
2\hbar\omega=\sqrt{c^2{\bf
p}^2+M_{\rm e}^2c^4}-\sqrt{c^2{\bf
p}^2+M_{\rm g}^2c^4}\;,
\end{equation}
where we have used the energy conservation and mass defect.

\section{Motion-induced shifts for atoms (ions) trapped in a confining
potential}
A third example concerns frequency shifts for
trapped atoms (ions) in an external confining potential. In
this case, the conservation of energy and momentum does not exist. Therefore, the above consideration for free atoms is not valid, and the implementation of the
mass defect concept requires other approaches, which are
developed below.

\begin{figure}[t]
\centerline{\scalebox{0.3}{\includegraphics{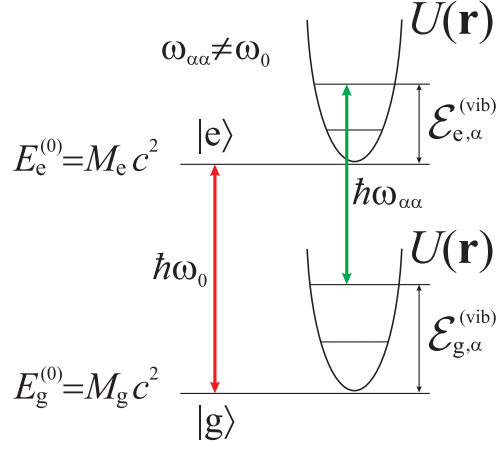}}}\caption{
Scheme of atomic transition $|{\rm g}\rangle\leftrightarrow |{\rm e}\rangle$. Also it is shown the quantization of energy levels on translational degrees of freedom in the confining potential $U({\bf r})$, where $\omega_{\alpha\alpha}\neq\omega_0$ due to the mass defect.} \label{Fig1}
\end{figure}

For simplicity, we will consider a stationary confining
potential $U({\bf r})$, which we take to be the same for both
states $|{\rm g}\rangle$ and $|{\rm e}\rangle$. Such a situation
can occur both for clocks based on neutral atoms in an optical
lattice and those based on trapped ions. In this case, we use the
standard formalism that quantizes the energy levels with
translational degrees of freedom:
\begin{equation}\label{E_vib}
\hat{H}_j|\Psi_{j,\alpha}({\bf r})\rangle={\cal E}^{{\rm (vib)}}_{j,\alpha}|\Psi_{j,\alpha}({\bf
r})\rangle\,,\;\; \hat{H}_j=\frac{\hat{\bf p}^2}{2M_j}+U({\bf
r})\,,
\end{equation}
where Hamiltonian $\hat{H}_j$ describes the translational motion of the particle in the $j$-th internal state $|j\rangle$ ($j={\rm g},{\rm e}$), wavefunction $|\Psi_{j,\alpha}({\bf r})\rangle$ corresponds to the $\alpha$-th vibrational level ($\alpha=0,1,2,...$) of the $j$-th internal state $|j\rangle$, and
${\bf r}$ is coordinate of atomic center-of-mass. Thus, taking
into account the translational motion, the atomic wave function is
described by the pair products $|j\rangle \otimes|\Psi_{j,\alpha}({\bf
r})\rangle$. Because of the mass defect ($M_{\rm e}\neq M_{\rm g}$),
the energy levels for the lower and upper states differ: ${\cal
E}^{{\rm (vib)}}_{{\rm g},\alpha}\neq {\cal E}^{{\rm (vib)}}_{{\rm e},\alpha}$. Consequently, the frequency
$\omega_{\alpha\alpha}$ between corresponding levels of the trapped particle
is different from the unperturbed frequency, $\omega_0$ (see Fig.~\ref{Fig1}), with a value $\Delta\omega$ (see Fig.~\ref{Fig1}):
\begin{equation}\label{DE_vib}
\Delta\omega_{\alpha\alpha}=\omega_{\alpha\alpha}-\omega_0=\left({\cal E}^{{\rm (vib)}}_{{\rm e},\alpha}-{\cal E}^{{\rm (vib)}}_{{\rm g},\alpha}\right)/\hbar\,.
\end{equation}
Let us now estimate this value. For this purpose, we write the Hamiltonian for upper state $\hat{H}_{\rm e}$ in the following form:
\begin{equation}\label{H_e}
\hat{H}_{\rm e}=\hat{H}_{\rm g}+\Delta \hat{H};\;
\Delta \hat{H}=\frac{\hat{\bf p}^2}{2M_{\rm e}}-\frac{\hat{\bf p}^2}{2M_{\rm g}}=\frac{M_{\rm g}-M_{\rm e}}{2M_{\rm g} M_{\rm e}}\hat{\bf p}^2=-\frac{\hbar\omega_0}{2c^2}\frac{\hat{\bf p}^2}{M_{\rm g} M_{\rm e}}\,,
\end{equation}
where the operator $\Delta \hat{H}$ can be considered as small perturbation. In this case, using standard perturbation theory, energy ${\cal E}^{{\rm (vib)}}_{{\rm e},\alpha}$ can be written as a series ${\cal E}^{{\rm (vib)}}_{{\rm e},\alpha}={\cal E}^{{\rm (vib)}}_{{\rm e},\alpha}(0)+\Delta{\cal E}^{{\rm (vib)}}_{{\rm e},\alpha}(1)+\Delta{\cal E}^{{\rm (vib)}}_{{\rm e},\alpha}(2)+...$, where ${\cal E}^{{\rm (vib)}}_{{\rm e},\alpha}(0)={\cal E}^{{\rm (vib)}}_{{\rm g},\alpha}$, and the first correction is determined as the average value $\Delta{\cal E}^{{\rm (vib)}}_{{\rm e},\alpha}(1)=\langle \Psi_{{\rm g},\alpha}|\Delta \hat{H}|\Psi_{{\rm g},\alpha}\rangle$. Using Eq.~(\ref{H_e}) and taking into account $M_{\rm e}\approx M_{\rm g}$, we obtain the following estimation of the relative value of shift (\ref{DE_vib}):
\begin{equation}\label{D_frac}
\frac{\Delta\omega_{\alpha\alpha}}{\omega_0}\approx-\frac{1}{2c^2}\frac{\langle \Psi_{{\rm g},\alpha}| \hat{\bf p}^2|\Psi_{{\rm g},\alpha}\rangle}{M_{\rm g}M_{\rm e}}
\approx-\frac{1}{2c^2}\frac{\langle \Psi_{{\rm g},\alpha}| \hat{\bf p}^2|\Psi_{{\rm g},\alpha}\rangle}{M^2_{\rm g}}\,.
\end{equation}
We note that this expression coincides with a well-known relativistic correction, which is the quadratic Doppler shift due to the time dilation effect for moving particle \cite{Chou_Sc_2010}. We see this correspondance if we consider the conventional explanation for this effect. In accordance with special relativity, the tick rate $\Delta t'$ in the moving (with velocity ${\bf v}$) coordinate system changes with respect to the tick rate $\Delta t$ in motionless (laboratory) coordinate system by the law: $\Delta t'=\Delta t\sqrt{1-{\bf v}^2/c^2}$. As a result,  an atomic oscillation with eigenfrequency $\omega_0$ is perceived by an external observer to be shifted to:
$\omega=\omega_0\sqrt{1-{\bf v}^2/c^2}$. In the nonrelativistic limit, (${\bf v}^2/c^2\ll 1$), we have:
$\omega\approx\omega_0[1-{\bf v}^2/(2c^2)]=\omega_0[1-({\bf p}/M)^2/(2c^2)]$ (where ${\bf p}$ is ���� momentum of particle). Then, if we take into account quantum considerations through the replacement ${\bf p}\rightarrow \hat{{\bf p}}=-i\hbar{\bf \nabla}$, we obtain the expression for frequency shift (\ref{D_frac}).

Thus, the conventional explanation formally requires the use of the proper
(internal) time of the atom. However, because of the wave nature of
quantum objects and the probabilistic interpretation of quantum
mechanics, a notion of the proper time (as some continuum) of
a quantum particle, spatially localized in a confined potential, seems unclear and even contradictory. In this context, an
approach, based on the mass defect concept, has not internal logical
contradictions and comes to the canonical quantum-mechanical
scheme (i.e., Hamiltonians, eigenvalues, eigenfunctions), which does not require the notion of
the proper time of quantum particle. Note also that due to the mass defect a general definition of the atomic velocity operator $\hat{{\bf v}}$ does not exists at all. Indeed, using the following quantum-relativistic formulas
\begin{equation}\label{V_j_oper}
\hat{{\bf p}}=\frac{M_j\hat{{\bf v}}_j}{\sqrt{1-\hat{{\bf v}}_j^2/c^2}}\;\Rightarrow \;\hat{{\bf v}}_j=\frac{\hat{{\bf p}}/M_j}{\sqrt{1+\hat{{\bf p}}^2/(M_j\,c)^2}}\,,
\end{equation}
we can rigorously determine only the particular velocity operator $\hat{{\bf v}}_j$, which is associated with only selected atomic energy state $|j\rangle$ (with the mass $M_j$).

To define more precisely the frequency shift for trapped atoms
(ions), we need to find the stationary states from the
Klein-Gordon equation (for $J_{\rm g}=0\rightarrow J_{\rm e}=0$
clock transition):
\begin{eqnarray}\label{E_vib_rel}
&&\{[E^{\rm (tot)}_j-U({\bf r})]^2-c^2\hat{\bf p}^2-M^2_j c^4\}|\Psi_j\rangle =0\,,  \nonumber \\
&& E^{\rm (tot)}_j=E^{(0)}_j+{\cal E}^{{\rm (vib)}}_j\,,\quad (j={\rm g,e}),
\end{eqnarray}
in the confining potential $U({\bf r})$. Note that the resulting
shift $\Delta\omega/\omega_0$ will not be equal to the well-known
phenomenological expression (see in
Ref.~\cite{Chou_Sc_2010}):
\begin{eqnarray}\label{phenom}
\frac{\Delta\omega}{\omega_0}=\frac{1}{\langle
(1-v_{||}/c)/\sqrt{1-{\bf v}^2/c^2}\,\rangle
}-1\,,\quad (v_{||}=({\bf n}\cdot {\bf v}))\,.
\end{eqnarray}
In the general case, the shift obtained from Eq.~(\ref{phenom}) will coincide with the {\em exact} shift obtained from Eq.~(\ref{E_vib_rel}) only in the
second order Doppler term.

In the case of dynamic confinement of an ion (with charge $eZ_i$) in the time-dependent electric potential $\phi(t,{\bf r})$ of an rf Paul trap (see in
\cite{Dehmelt_1968,Ludlow_2015}):
\begin{equation}\label{Ut}
\phi(t,{\bf r})=V_{\rm dc\,}a({\bf r})+V_{\rm rf\,}b({\bf r})\cos(ft)\,,
\end{equation}
where $a({\bf r})$ and $b({\bf r})$ describe spatial dependencies, the mass defect concept is also applicable and can explain, in
particular, the frequency shift, which is usually interpreted as
the micromotion shift. Indeed, in the case of relatively high rf
frequency $f$ in the Paul trap, there is the well-working approximation of
the so-called pseudopotential, which can be expressed for $j$-th internal state $|j\rangle$ (in
context of the mass defect) as the following sum:
\begin{equation}\label{pseudo}
U_{\rm pseudo}^{(j)}({\bf r})=eZ_iV_{\rm dc\,}a({\bf r})+\frac{W_{\rm rf}({\bf r})}{M_j}\,,\;\; W_{\rm rf}({\bf r})=\frac{e^2Z_i^2V_{\rm rf}^2}{4f^2}({\bf \vec{\nabla}} b({\bf r}))^2,
\end{equation}
where the first term is connected with the static electric
potential, the second term is produced by the rf oscillating
potential [see in Eq.~(\ref{Ut})] and it is related to the
micromotion (${\bf \vec{\nabla}}$ is the gradient operator). Note that the approach of pseudopotential Eq.~(\ref{pseudo}) is valid for description of the lower vibrational levels: ${\cal E}^{{\rm (vib)}}_{j,\alpha}\ll \hbar f$.

As we see, the second term in Eq.~(\ref{pseudo}) is
mass-dependent. In this case, the total Hamiltonian for upper
state, $\hat{H}_{\rm e}^{\rm (eff)}=\hat{\bf
p}^2/(2M_{\rm e})+U_{\rm pseudo}^{\rm (e)}({\bf r})$, can be
written in the following form:
\begin{eqnarray}\label{He_pseudo}
\hat{H}_{\rm e}^{\rm (eff)}&=&\hat{H}_{\rm g}^{\rm (eff)}+\Delta \hat{H}^{\rm (eff)},\\
\Delta \hat{H}^{\rm (eff)}&=&\left(\frac{1}{M_{\rm e}}-\frac{1}{M_{\rm g}}\right)\left[\frac{\hat{\bf p}^2}{2}+W_{\rm rf}({\bf r})\right]
=-\frac{\hbar\omega_0}{M_{\rm g}
M_{\rm e}c^2}\left[\frac{\hat{\bf p}^2}{2}+W_{\rm rf}({\bf
r})\right],\nonumber
\end{eqnarray}
where the operator $\Delta \hat{H}^{\rm (eff)}$ can be considered a small perturbation. Consequently, we have the frequency shift:
\begin{eqnarray}\label{D_frac_pseudo}
\frac{\Delta\omega_{\alpha\alpha}}{\omega_0}&\approx&-\frac{1}{M_{\rm g} M_{\rm e}c^2}\left[\frac{\langle \Psi_{{\rm g},\alpha}| \hat{\bf p}^2|\Psi_{{\rm g},\alpha}\rangle}{2}+\langle \Psi_{{\rm g},\alpha}| W_{\rm rf}({\bf r})|\Psi_{{\rm g},\alpha}\rangle\right]\nonumber \\
&\approx&-\frac{1}{2c^2}\frac{\langle \Psi_{{\rm g},\alpha}| \hat{\bf
p}^2|\Psi_{{\rm g},\alpha}\rangle}{M_{\rm g}^2}-\frac{\langle\Psi_{{\rm g},\alpha}|
W_{\rm rf}({\bf r})|\Psi_{{\rm g},\alpha}\rangle}{M_{\rm g}^2c^2}\,,
\end{eqnarray}
where the first contribution coincides with Eq.~(\ref{D_frac}),
and which can be considered as secular-motion-induced shift. The
second term in Eq.~(\ref{D_frac_pseudo}) can be interpreted as
micromotion shift. Note that both these shifts have comparable
values, in the general case.

In the case of a purely rf Paul trap with $V_{\rm dc}=0$ in Eq.~(\ref{pseudo}), we have the following relationships for Hamiltonians, vibrational energies and eigenfunctions:
\begin{equation}\label{purely_rf}
\hat{H}_{\rm e}^{\rm (eff)}=\frac{M_{\rm g}}{M_{\rm e}}\hat{H}_{\rm g}^{\rm (eff)},\; {\cal E}^{{\rm (vib)}}_{{\rm e},\alpha}=\frac{M_{\rm g}}{M_{\rm e}}\,{\cal E}^{{\rm (vib)}}_{{\rm g},\alpha},\;\Psi_{{\rm e},\alpha}({\bf r})=\Psi_{{\rm g},\alpha}({\bf r})\,,
\end{equation}
leading to the total fractional frequency shift:
\begin{equation}\label{sh_purely_rf}
\frac{\Delta\omega_{\alpha\alpha}}{\omega_0}=-\frac{{\cal E}^{{\rm (vib)}}_{{\rm g},\alpha}}{M_{\rm e}c^2}\approx -\frac{{\cal E}^{{\rm (vib)}}_{{\rm g},\alpha}}{M_{\rm g}c^2}\,,
\end{equation}
which contains both secular-motion and micromotion contributions [see Eq.~(\ref{D_frac_pseudo})].

Note that in a similar way we can consider the mass defect effects in the case of several ions confined in the same trap.

\section{Previously unconsidered field-induced shifts for trapped
ions}
Besides the reinterpretation of some well-known shifts,
the mass defect concept predicts additional contributions for
field-induced shifts that have not been previously discussed in the
scientific literature. We emphasize that these additional shifts
are associated with translational degrees of freedom and they
vanish if we will not take into account the mass defect.

We will consider a trapped ion (with charge $eZ_i$)  in the presence of additional weak electric field with potential $\varphi_{\rm add}({\bf r})$, which describes all controlled and uncontrolled fields except the trapping potential $U({\bf r})$ from previous Section 3. Let us show how an additional potential $\varphi_{\rm add}({\bf r})$ will perturb the vibrational structure formed by $U({\bf r})$. For this purpose we will use vibrational eigenfunctions $|\Psi_{{\rm g},\alpha}({\bf r})\rangle$ and $|\Psi_{{\rm e},\alpha}({\bf r})\rangle$, which describe a spatial localization of the ion in the internal states $|{\rm g}\rangle$ and $|{\rm e}\rangle$ due to the trap potential $U({\bf r})$, as a basis for perturbation theory on the small additional interaction $U_{\rm add}({\bf r})=eZ_i\varphi_{\rm add}({\bf r})$. This addition describes an interaction of the point charge $eZ_i$ with potential $\varphi_{\rm add}({\bf r})$.

The center of the ion localization ${\bf r}_0$ in the trap is determined by averaging: ${\bf r}_0=\langle\Psi_{j,\alpha}|{\bf r}|\Psi_{j,\alpha}\rangle$. In this case, we will use the following Taylor series of the additional operator $U_{\rm add}({\bf r})$ at the point of ion localization ${\bf r}_0$:
\begin{eqnarray}\label{phi_add}
&&U_{\rm add}({\bf r})= eZ_{i\,}\varphi_{\rm add}({\bf r})=eZ_{i\,}\varphi_{\rm add}({\bf r}_0+({\bf r}-{\bf r}_0))\nonumber \\
&& =eZ_i\left[\varphi_{\rm add}({\bf r}_0)-\left(({\bf r}-{\bf r}_0)\cdot{\bf E}_{\rm add}({\bf r}_0)\right)+\sum_{q,q'=1}^{3}Q_{qq'}^{\rm (norm)}W_{q'q}^{\rm (add)}({\bf r}_0)+... \right]\nonumber\\
&& =eZ_{i\,}\varphi_{\rm add}({\bf r}_0)-\left(\hat{{\bf d}}^{\rm (cloud)}\cdot{\bf E}_{\rm add}({\bf r}_0)\right)+{\rm Tr}\left\{\hat{Q}^{\rm (cloud)}\hat{W}^{\rm (add)}({\bf r}_0)\right\}+...\,,
\end{eqnarray}
where ${\bf E}_{\rm add}({\bf r}_0)=-\nabla_{\bf r} \varphi_{\rm add}({\bf r})|_{{\bf r}_0}$ and $\hat{W}^{\rm (add)}({\bf r}_0)=W_{q'q}^{\rm (add)}({\bf r}_0)$ are electric vector and electric tensor, respectively, at the point ${\bf r}_0$. Operators $\hat{{\bf d}}^{\rm (cloud)}$ and $\hat{Q}^{\rm (cloud)}$ describe {\em mesoscopic} dipole and quadrupole moments, respectively, of ion cloud:
\begin{eqnarray}\label{d_Q}
&&\hat{{\bf d}}^{\rm (cloud)}=eZ_i({\bf r}-{\bf r}_0)\,, \nonumber \\
&&\hat{Q}^{\rm (cloud)}=Q_{qq'}^{\rm (cloud)}=eZ_i\left\{3({\bf r}-{\bf r}_0)_q({\bf r}-{\bf r}_0)_{q'}-\delta_{qq'}|{\bf r}-{\bf r}_0|^2 \right\},
\end{eqnarray}
where $({\bf r}-{\bf r}_0)_q$ is $q$-th Cartesian coordinate of the vector $({\bf r}-{\bf r}_0)$. The first term in the last line of Eq.~(\ref{phi_add}) is unimportant, and we can use the following condition: $\varphi_{\rm add}({\bf r}_0)=0$ (we assume, for simplicity, that ${\bf r}_0$ is the same for all states $|\Psi_{j,\alpha}({\bf r})\rangle$). The obtained results can be reproduced in the general case of four-vector \{$\varphi_{\rm add}(t,{\bf r}), {\bf A}_{\rm add}(t,{\bf r})$\}. Note also that the approach developed above is suitable if the size of spatial nonuniformity (e.g., wavelength $\lambda$) of the additional field is much more than the size of ion cloud:
\begin{equation}\label{cloud}
R_{\rm cloud}=\sqrt{\langle\Psi_{j,\alpha}|({\bf r}-{\bf r}_0)^2|\Psi_{j,\alpha}\rangle} \;,
\end{equation}
which is a typical size of the wavefunctions $|\Psi_{j,\alpha}({\bf r})\rangle$.

Thus, we have shown that apart from well-known {\em electronic} dipole and quadrupole moments a trapped ion has additional mesoscopic dipole and qudrupole moments, which describe an interaction of the ion cloud (formed by the trapping potential $U({\bf r})$) with weak external fields. This interaction involves only translational degrees of freedom ({\bf r}) and it leads to a perturbation (i.e., frequency shifts) of the vibrational structure in trapped ions. However, for manifestation of these shifts in atomic clocks we need to take into account the mass defect concept, because without mass defect these shifts will not lead to a shift of the clock transitions.

As a first example, let us consider the first-order shift, described by diagonal elements, $\Delta_{j,\alpha}^{(1)}=\langle\Psi_{j,\alpha}|U_{\rm add}({\bf r})|\Psi_{j,\alpha}\rangle/\hbar$. Because of $\langle\Psi_{j,\alpha}|\hat{{\bf d}}^{\rm (cloud)}|\Psi_{j,\alpha}\rangle =0$ (see Eq.~(\ref{d_Q})), we obtain:
\begin{eqnarray}\label{Delta_1}
&&\Delta_{j,\alpha}^{(1)}={\rm Tr}\left\{\langle\hat{Q}^{\rm (cloud)}\rangle_{j,\alpha}\hat{W}^{\rm (add)}({\bf r}_0)\right\}/\hbar \,,\nonumber \\
&&\langle\hat{Q}^{\rm (cloud)}\rangle_{j,\alpha}=\langle\Psi_{j,\alpha}|\hat{Q}^{\rm (cloud)}|\Psi_{j,\alpha}\rangle,\quad (j={\rm g,e}).
\end{eqnarray}
Consequently, we have the following residual shift of the clock transition $|{\rm
g}\rangle\leftrightarrow |{\rm e}\rangle$:
\begin{equation}\label{res_1}
\bar{\delta}_{\alpha}^{(1)}=\Delta_{{\rm e},\alpha}^{(1)}-\Delta_{{\rm g},\alpha}^{(1)}={\rm Tr}\left\{\Delta\hat{Q}^{\rm (m-def)}_{\alpha}\hat{W}^{\rm (add)}({\bf r}_0)\right\}/\hbar \,,
\end{equation}
where $\Delta\hat{Q}^{\rm (m-def)}_{\alpha}$ is a residual qudrupole moment:
\begin{equation}\label{Q_res}
\Delta\hat{Q}^{\rm (m-def)}_{\alpha}=\langle\Psi_{{\rm e},\alpha}|\hat{Q}^{\rm (cloud)}|\Psi_{{\rm e},\alpha}\rangle -\langle\Psi_{{\rm g},\alpha}|\hat{Q}^{\rm (cloud)}|\Psi_{{\rm g},\alpha}\rangle ,
\end{equation}
which can be nonzero, because $|\Psi_{{\rm g},\alpha}({\bf r})\rangle\neq |\Psi_{{\rm e},\alpha}(\bf r)\rangle$, in the general case, due to the mass defect. Let us estimate an order of $\Delta\hat{Q}^{\rm (m-def)}$:
\begin{equation}\label{DQ_vel}
|\Delta \hat{Q}^{\rm (m-def)}_{\alpha}|\sim |\langle \Psi_{{\rm g},\alpha}|\hat{Q}^{\rm (cloud)}|\Psi_{{\rm g},\alpha}\rangle|\frac{\hbar \omega_0}{Mc^2}\sim eZ_i  R_{\rm cloud}^2\frac{\hbar \omega_0}{Mc^2}\,,
\end{equation}
where $R_{\rm cloud}$ is a size of ion cloud (see in Eq.~(\ref{cloud})), and $M\approx
M_{\rm g,e}$. Though the expression (\ref{DQ_vel}) contains a
very small multiplier, $\hbar\omega_0/Mc^2\ll 1$, the size
of ion localization $R_{\rm cloud}$ significantly exceeds the Bohr radius
$a_0$. Indeed, $R_{\rm cloud}\sim 10^2$--$10^3a_0$ even for the
deeply-cooled ion to the lowest vibrational level in the confined
potential $U({\bf r})$ (i.e., for the quantum limit of cooling),
and $R_{\rm cloud}\sim 10^4a_0$ for the upper vibrational states, which are
populated if the ion is laser-cooled to the usual so-called
Doppler temperature (mK range). As a result, the quadrupole shift
of the clock transition, modified by the  mass defect, can be
metrologically significant for modern and future optical frequency
standards. For example, let us consider an atomic clock based on
the transition $^{1}{\rm S}_0$$\rightarrow$$^3{\rm P}_0$ in the ion
$^{27}$Al+ \cite{Chou_2010}. Because of the zero electronic
angular momentum for the clock transition,
$J_{\rm g}=J_{\rm e}=0$, the quadrupole moment, associated with
internal degrees of freedom, is very small ($|\Delta Q|\sim
10^{-6}ea^2_0$, see Ref.~\cite{Beloy_2017}). However, on the
basis of formula (\ref{DQ_vel}), we estimate $|\Delta
\hat{Q}^{\rm (m-def)}|$$\sim$ 2$\times (10^{-2}$--$10^{-6})ea^2_0$
for $R_{\rm cloud}\sim$~$10^2$--$10^4a_0$. As another example, let us consider
the so-called nuclear clock, based on the intranuclear transition
in $^{229}{\rm Th}^{3+}$ \cite{Campbell_2012}. In
Ref.~\cite{Campbell_2012}, the quadrupole moment (associated with internal degrees of freedom) for the
clock transition was estimated to be: $|\Delta Q|\sim
10^{-5}ea^2_0$. Using now our formula (\ref{DQ_vel}), we find
$|\Delta \hat{Q}^{\rm (m-def)}|$$\sim$ $10^{-2}$--$10^{-6}ea^2_0$ for $R_{\rm cloud}\sim$~$10^2$--$10^4a_0$.
Such values of the quadrupole moment may be important for atomic
clocks with fractional uncertainty at the level of
$10^{-18}$--$10^{-19}$. However, for more accurate estimations, it
is necessary to know the wave functions on the translational
degrees of freedom $|\Psi_{j,\alpha}({\bf r})\rangle$ ($j={\rm g,e}$). In particular, this quadrupole shift is absent for
spherically-symmetrical states $|\Psi_{j,\alpha}({\bf r})\rangle=|\Psi_{j,\alpha}(|{\bf r}-{\bf r}_0|)\rangle$, when $\langle\Psi_{{\rm
g},\alpha}|\hat{Q}^{\rm (cloud)}|\Psi_{{\rm g},\alpha}\rangle =0$ and $\langle\Psi_{{\rm
e},\alpha}|\hat{Q}^{\rm (cloud)}|\Psi_{{\rm
e},\alpha}\rangle =0$. Also the residual
quadruple moment (\ref{Q_res}) vanishes for a purely rf Paul trap,
because of $|\Psi_{{\rm g},\alpha}({\bf r})\rangle =|\Psi_{{\rm
e},\alpha}({\bf r})\rangle$ [see Eq.~(\ref{purely_rf})].

To see another manifestation of mass defect, let us consider a previously unknown
contribution to the ac-Stark shift in the presence of weak external
low-frequency field, ${\bf E}(t)=({\bf E}_s^{} e^{-i\nu t}+{\bf E}_s^{*} e^{i\nu t})$. Here we will
investigate only shifts of the quantum levels ${\cal E}^{{\rm (vib)}}_{{\rm g},\alpha}$ and ${\cal E}^{{\rm (vib)}}_{{\rm e},\alpha}$ on the translational degrees
of freedom of the trapped ion (see
Fig.~\ref{Fig1}). This shift exists, because some nondiagonal elements of the dipole moment between two different vibrational states $\alpha$ and $\alpha'$ can be already nonzero:  ${\bf d}_{\alpha\alpha'}^{\,(j)}=\langle\Psi_{j,\alpha}|\hat{{\bf d}}^{\rm (cloud)}|\Psi_{j,\alpha'}\rangle \neq0$ ($\alpha\neq\alpha'$). In this case, the expression for
the dynamic shift of the $\alpha$-th vibrational level ${\cal
E}_{j,\alpha}$ in the internal state $|j\rangle$ has the following standard form:
\begin{equation}\label{ac_Stark}
\Delta^{(2)}_{j,\alpha}=\frac{1}{\hbar}\sum_{\alpha'}\left[\frac{|{\bf E}_s^{}\cdot{\bf d}_{\alpha'\alpha}^{\,(j)}|^2}{\hbar\nu -{\cal E}_{j,\alpha'}+{\cal E}_{j,\alpha}}+
\frac{|{\bf E}_s^{*}\cdot{\bf d}_{\alpha'\alpha}^{\,(j)}|^2}{-\hbar\nu -{\cal E}_{j,\alpha'}+{\cal E}_{j,\alpha}}\right] ,\;\; (j={\rm g,e})\,.
\end{equation}
Note that the square of
the ion-cloud dipole moment has the order-of-magnitude of $|{\bf d}_{\alpha\alpha'}^{\,(j)}|^2\sim e^2Z_i^2R^2_{\rm cloud}$ (where $R_{\rm cloud}$ is typical size of localization of wavefunctions $|\Psi_{j,\alpha}({\bf r})\rangle$, see above), which is many orders greater than the square of the atomic dipole moment
$e^2a_0^2$, because $R_{\rm cloud}\gg a_0$. Thus, the vibrational levels of a trapped ion are very sensitive even to the weak external fields. However, because $\Delta^{(2)}_{{\rm e},\alpha}$ and
$\Delta^{(2)}_{{\rm g},\alpha}$ are slightly different only due to the mass defect, we find and estimate the differential shift of the clock transition $|{\rm g}\rangle\leftrightarrow |{\rm e}\rangle$:
 \begin{equation}\label{ac_diff}
\bar{\delta}^{(2)}_{\alpha}=\Delta^{(2)}_{{\rm e},\alpha}-\Delta^{(2)}_{{\rm g},\alpha}\,,\quad |\bar{\delta}^{(2)}_{\alpha}|\sim
|\Delta^{(2)}_{{\rm g},\alpha}| \frac{\hbar\omega_0}{Mc^2}\,,
\end{equation}
which can exceed the ac-Stark shift connected with low-frequency
polarizability, which is associated with internal (electronic) degrees of freedom.

In a similar way, we can find new contributions to the black body
radiation (BBR) shift, linear and quadratic Zeeman shifts, and so
on. Note that we have formulated here only a general qualitative
approach to the description of previously unknown shifts, which
require more detailed consideration in the further investigations.

\section*{Conclusion}
We have considered some manifestations of the mass defect in atomic clocks. As a result, some well-known systematic shifts, previously interpreted as the time dilation effects in the frame of special and general relativity theories, can be considered as a consequence of the mass defect in quantum mechanics. In particular, we have derived previously unknown exact quantum-relativistic formulas (\ref{omega_abs_p})-(\ref{omega_rad}) and (\ref{E_2_ph}) for one-photon and two-photon transitions for free atoms. Also we have obtained previously unknown analytical expression for micromotion shift (\ref{D_frac_pseudo}) for ion trapped in rf Paul trap. Furthermore, our approach has predicted a series of previously unknown field-induced shifts for ion clocks. These results are important for high-precision atomic clocks and can be interesting for theoretical quantum physics.

We thank C. W. Oates, K. Beloy, E. Peik, A. Derevianko, and N. Kolachevsky
for helpful discussions and comments.

\end{document}